\documentclass[11pt]{article}
\usepackage{authblk}
\usepackage[a4paper, margin=1in]{geometry}
\usepackage{amsmath, amssymb}
\usepackage{graphicx}

\usepackage{braket}
\usepackage{booktabs}

\title{Relativistic Quantum Simulation of Hydrogen Sulfide for Hydrogen Energy via Hybrid Quantum-Classical Algorithms}

\author[1]{Chi-Chuan Hwang}
\author[2, 3, 4]{Cheng-Fang Su\thanks{Corresponding author. Email: scf1204@nycu.edu.tw}}
\author[1]{Jung-Fan Yang}

\affil[1]{Department of Engineering Science, National Cheng Kung University, Tainan City, 701, Taiwan}

\affil[2]{Department of Applied Mathematics, National Yang Ming Chiao Tung University, Hsinchu City, 30010, Taiwan}

\affil[3]{Undergraduate Degree Program of Systems Engineering and Technology, National Yang Ming Chiao Tung University, Hsinchu City, 30010, Taiwan}

\affil[4]{Chung Cheng Institute of Technology, National Defense University, Taoyuan City, 33551, Taiwan}

\usepackage{hyperref}
\usepackage{caption}
\usepackage{cite}
\usepackage{bm}
\usepackage{float}
\usepackage{color}

\hypersetup{
  colorlinks=true,
  linkcolor=blue,
  citecolor=blue,
  urlcolor=blue
}

\begin{document}
\maketitle

\begin{abstract}
We present a hybrid quantum-classical computational framework for simulating relativistic quantum chemical systems using the variational quantum eigensolver (VQE). Our approach integrates relativistic electron integrals derived from the Dirac–Coulomb Hamiltonian with quantum circuit models constructed via Jordan–Wigner transformation. By employing both heuristic and chemistry-inspired ansätze, we compute ground-state energies and potential energy surfaces (PES) for benchmark molecules including H$_2$, H$_2$O, and H$_2$S. Active-space reduction and Hartree–Fock embedding techniques are used to minimize the number of qubits and Pauli terms, enabling efficient simulation on near-term quantum devices.

Our results demonstrate that VQE can reproduce classical full-configuration-interaction energies with millihartree-level accuracy, even under relativistic Hamiltonians. We observe consistent relativistic energy shifts and orbital contraction effects, particularly for the sulfur-containing H$_2$S molecule. The constructed PES allow for equilibrium geometry prediction and activation energy analysis. This work highlights the feasibility and scalability of applying quantum algorithms to relativistic electronic structure problems, paving the way for future simulations of heavier-element systems in quantum chemistry.
\end{abstract}

\section{Introduction}
Quantum computing has emerged as a transformative paradigm for solving computationally intractable problems in physics, chemistry, and materials science. Since Feynman's foundational proposal to simulate quantum systems with quantum devices~\cite{Feynman1982}, numerous algorithms have been developed to address the limitations of classical methods, particularly in electronic structure calculations~\cite{AspuruGuzik2005, Peruzzo2014}. Recent demonstrations of quantum advantage~\cite{Google2019} and the increasing accessibility of cloud-based quantum platforms (e.g., IBM Quantum~\cite{IBMQ}) have accelerated the deployment of hybrid quantum-classical algorithms for practical applications.

In computational chemistry, solving the electronic Schrödinger equation to obtain molecular energies and wavefunctions remains a central challenge. While classical algorithms such as full configuration interaction (FCI) are accurate, their exponential scaling renders them impractical for all but the smallest molecules. The variational quantum eigensolver (VQE)~\cite{Peruzzo2014} has emerged as a leading candidate for near-term quantum devices, leveraging parameterized quantum circuits and classical optimization to approximate ground-state energies efficiently.

While VQE has been widely adopted in quantum chemistry due to its suitability for noisy intermediate-scale quantum (NISQ) devices, most existing studies focus on light-element molecules using non-relativistic Hamiltonians. For example, Peruzzo et al.~\cite{Peruzzo2014} demonstrated early VQE implementations on small molecules, while Kandala et al.~\cite{Kandala2017} employed hardware-efficient ansätze to simulate LiH and BeH$_2$. McClean et al.~\cite{McClean2016} analyzed the optimization landscape of VQE and highlighted the challenges of barren plateaus in high-dimensional variational spaces.

Recent efforts have begun addressing the limitations of standard optimization routines in VQE. For instance, Fitzek et al.~\cite{Fitzek2024} proposed the \textit{qBang} algorithm, which combines Broyden-based metric estimation with momentum-based updates to improve convergence in flat energy landscapes. While their work enhances VQE optimization performance, it remains focused on generic systems and non-relativistic chemistry.

In parallel, Rice et al.~\cite{Rice2020} explored the application of VQE to simulate dissociation curves and dipole moments of lithium–sulfur battery molecules, including H$_2$S and LiSH. Their simulations, however, were carried out using classical quantum simulators and standard (non-relativistic) Hamiltonians, without incorporating relativistic corrections.

Cao et al.~\cite{Cao2019} provided a comprehensive review of quantum computing in quantum chemistry, highlighting the importance of variational algorithms and outlining the challenges of integrating relativistic effects. Although they emphasize the theoretical potential for such developments, practical implementations remain scarce.

Relativistic effects—including mass–velocity corrections, spin–orbit coupling, and orbital contraction—play a critical role in accurately modeling molecules involving elements beyond the second period~\cite{Dirac1928}. The Dirac–Coulomb Hamiltonian provides a first-principles relativistic formulation but introduces technical challenges such as negative-energy solutions and spinor representations. Despite advances in classical relativistic quantum chemistry~\cite{DIRACprogram}, the integration of relativistic physics into quantum variational algorithms remains underexplored.

This work addresses that gap. We present a hybrid quantum-classical framework that integrates relativistic quantum chemistry with VQE, using one- and two-electron integrals computed from the Dirac–Coulomb Hamiltonian. We simulate small molecules—including H$_2$, H$_2$O, and H$_2$S—on a quantum backend, with particular emphasis on sulfur-containing species where relativistic corrections become non-negligible. Our approach employs Jordan–Wigner mapping, orbital embedding, and ansatz optimization to reduce circuit depth and enable efficient simulation.

Compared with prior works, our study provides the first fully-integrated VQE simulation on relativistic systems using explicitly constructed Dirac-based Hamiltonians. It complements recent advancements in optimization~\cite{Fitzek2024}, broadens the scope of molecular simulation beyond non-relativistic domains~\cite{Rice2020}, and concretely implements the future directions envisioned in major reviews~\cite{Cao2019}.

This study expands on previous thesis work by J.-F. Yang~\cite{Yang2022}, originally conducted at National Cheng Kung University. The current article revises the simulation pipeline, extends the theoretical framework to a wider class of molecules, and provides a comprehensive comparison between relativistic and non-relativistic treatments under the variational quantum eigensolver approach.

The remainder of this paper is structured as follows: Section~2 presents the theoretical foundation, including relativistic Hamiltonians and VQE components. Section~3 outlines the computational workflow and circuit construction. Section~4 details simulation results and discusses relativistic effects. Section~5 concludes with a summary and outlook.

\section{Theoretical Framework}
This section outlines the theoretical foundations upon which our study is based, particularly focusing on applying relativistic quantum chemistry to the electronic structure of the hydrogen sulfide (H$_2$S) molecule. We begin by introducing the relativistic treatment of molecular systems, followed by describing how these concepts are encoded and simulated on hybrid quantum-classical computing architectures.

\subsection{Relativistic Electronic Structure Theory}
Accurate modeling of molecular systems containing heavy elements requires the explicit inclusion of relativistic effects in electronic structure calculations. In particular, for molecules involving atoms such as sulfur, relativistic contributions—especially scalar relativistic corrections and spin--orbit coupling—can significantly influence orbital energies, electron densities, and molecular properties.

A first-principles relativistic description of the electronic structure is provided by the Dirac--Coulomb Hamiltonian. For a single electron in the field of a point nucleus, the Dirac equation is given by
\begin{equation}
    \hat{H}_D = c \, \boldsymbol{\alpha} \cdot \mathbf{p} + \beta m c^2 + V(\mathbf{r}),
\end{equation}
where $c$ denotes the speed of light, $\boldsymbol{\alpha}$ and $\beta$ are the Dirac matrices, $\mathbf{p}$ is the momentum operator, and $V(\mathbf{r})$ represents the nuclear Coulomb potential. Note that $\boldsymbol{\alpha}$ represents a three-component vector of $4 \times 4$ matrices $(\alpha_x, \alpha_y, \alpha_z)$, and is written in bold to indicate its vectorial character. In contrast, $\beta$ is a single Dirac matrix associated with the mass term and is not boldfaced.

For an $N$-electron system, the Dirac--Coulomb Hamiltonian takes the form
\begin{equation}
    \hat{H}_{\text{DC}} = \sum_i \left[ c \, \boldsymbol{\alpha}_i \cdot \mathbf{p}_i + \beta_i m c^2 + V(\mathbf{r}_i) \right] + \sum_{i<j} \frac{1}{r_{ij}},
\end{equation}
where the second term accounts for electron--electron Coulomb repulsion. This four-component formalism introduces spinor wavefunctions and incorporates relativistic effects in a nonperturbative manner.

In practice, relativistic molecular integrals are evaluated using quantum chemistry software packages such as DIRAC~\cite{DIRACprogram}, which perform mean-field calculations based on the Dirac--Hartree--Fock (DHF) approximation. The resulting one-electron and two-electron integrals, expressed in a relativistic basis, are subsequently transformed into a second-quantized fermionic Hamiltonian suitable for quantum simulation.

\subsection{Variational Quantum Eigensolver (VQE)}
The variational quantum eigensolver (VQE) is a hybrid quantum-classical algorithm designed to approximate the ground-state energy of a quantum system. It is based on the variational principle, which guarantees that the expectation value of any trial wavefunction provides an upper bound to the true ground-state energy:
\begin{equation}
    E_0 \leq \bra{\psi(\bm{\theta})} \hat{H} \ket{\psi(\bm{\theta})}.
\end{equation}
Here, $\ket{\psi(\bm{\theta})}$ is a parameterized quantum state constructed by a quantum circuit with variational parameters $\bm{\theta}$. The algorithm proceeds iteratively: a quantum device prepares the state and estimates expectation values of the Hamiltonian components, while a classical optimizer updates the parameters $\bm{\theta}$ to minimize the energy.

To perform VQE for quantum chemical systems, the fermionic Hamiltonian is mapped to a qubit Hamiltonian via encodings such as the Jordan–Wigner transformation~\cite{Jordan1928}, which preserves fermionic anti-commutation relations. The resulting Hamiltonian consists of sums of weighted Pauli strings, which are amenable to measurement on quantum devices.

Two classes of ansätze are typically used to construct the trial wavefunction:
(1) hardware-efficient ansätze, which are designed to be compatible with specific quantum hardware architectures and consist of shallow layers of parameterized gates and entanglers; and
(2) chemistry-inspired ansätze, such as the unitary coupled-cluster singles and doubles (UCCSD), which directly encode electron excitation operators from a reference Hartree–Fock state.

The total energy is reconstructed by evaluating the weighted average of measured Pauli strings. Multiple measurements per term are required due to the probabilistic nature of quantum measurement, leading to a measurement overhead that grows with Hamiltonian complexity.

Despite these challenges, VQE remains a promising algorithm for quantum chemistry simulations on near-term devices. In this work, we apply VQE to relativistic Hamiltonians derived from the Dirac–Coulomb framework, enabling us to study relativistic energy corrections and potential energy surfaces for heavy-element systems such as hydrogen sulfide.


\section{Computational Methodology}
This section outlines the computational strategy employed to simulate the electronic structure of hydrogen sulfide using a hybrid quantum-classical framework. Building upon the relativistic formulation established in the previous section, we translate the Dirac--Coulomb-based molecular Hamiltonian into a representation amenable to quantum simulation. This involves a sequence of classical preprocessing steps, quantum circuit design, and iterative variational procedures that collectively enable ground-state energy estimation on near-term quantum devices.

\subsection{Overview of Simulation Workflow}
The simulation workflow consists of four stages: (1) relativistic molecular integral generation; (2) fermion-to-qubit encoding; (3) quantum circuit construction; and (4) hybrid variational optimization. The process begins with relativistic Dirac–Coulomb integrals computed using the DIRAC quantum chemistry package. These integrals are used to construct a second-quantized Hamiltonian, which is then mapped to a qubit Hamiltonian using the Jordan–Wigner transformation. An initial Hartree–Fock state is prepared and evolved under a parameterized ansatz circuit. Energy expectation values are measured and minimized via a classical optimizer to obtain the ground-state energy.

\subsection{Qubit Mapping via Jordan--Wigner Transform}
To simulate molecular systems on quantum hardware, the fermionic creation and annihilation operators must be encoded in terms of Pauli matrices. We employ the Jordan–Wigner (JW) transformation, which maps each fermionic operator $a_j^\dagger$, $a_j$ onto qubit operators while preserving the anti-commutation relations:
\begin{align}
    a_j^\dagger &= \frac{1}{2}(X_j - iY_j) Z_0 Z_1 \cdots Z_{j-1}, \\
    a_j &= \frac{1}{2}(X_j + iY_j) Z_0 Z_1 \cdots Z_{j-1}.
\end{align}
This transformation results in qubit Hamiltonians represented as linear combinations of Pauli strings. Although JW encoding introduces non-locality scaling linearly with qubit index, its simplicity makes it particularly suitable for compact molecules such as H$_2$S with a modest number of orbitals.

\subsection{Initial State Preparation}
The reference state used in this study is the Hartree–Fock (HF) state. This state corresponds to occupying the lowest-energy spin orbitals as determined by mean-field calculations. The HF state is implemented on a quantum device by applying a sequence of Pauli-$X$ gates to the corresponding qubits. For example, if qubits 0, 1, and 3 are to be occupied, the HF state is constructed via:
\[
\ket{\psi_{\mathrm{HF}}} = X_0 X_1 X_3 \ket{0}^{\otimes n}.
\]
The HF state serves as the starting point for the ansatz-based evolution.

\begin{figure}[H]
    \centering
    \includegraphics[width=0.1\textwidth]{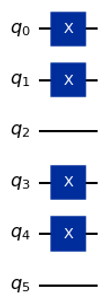}
    \caption{Quantum circuit for the initial state of H$_2$S after orbital reduction (adapted from~\cite{Yang2022}).}
    \label{fig:H2S_initial_state}
\end{figure}

\subsection{Ansatz Design: Heuristic and UCCSD}
We examine two classes of ansätze in this work. The first is a hardware-efficient ansatz, composed of alternating layers of single-qubit rotations and entangling gates, designed to minimize circuit depth. The second is a unitary coupled-cluster singles and doubles (UCCSD) ansatz, derived from classical quantum chemistry. The UCCSD ansatz takes the form:
\[
\ket{\psi_{\mathrm{UCCSD}}} = e^{T - T^\dagger} \ket{\psi_{\mathrm{HF}}},
\]
where $T$ contains single and double excitation operators. This ansatz, while more expressive, results in deeper circuits and requires Trotterization to implement efficiently.

\begin{figure}[H]
    \centering
    \includegraphics[width=0.8\textwidth]{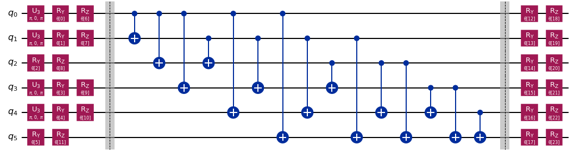}
    \caption{Heuristic ansatz circuit for H$_2$S. Composed of entangled layers and parameterized single-qubit rotations (adapted from~\cite{Yang2022}).}
    \label{fig:heuristic_circuit}
\end{figure}

\begin{figure}[H]
    \centering
    \includegraphics[width=0.9\textwidth]{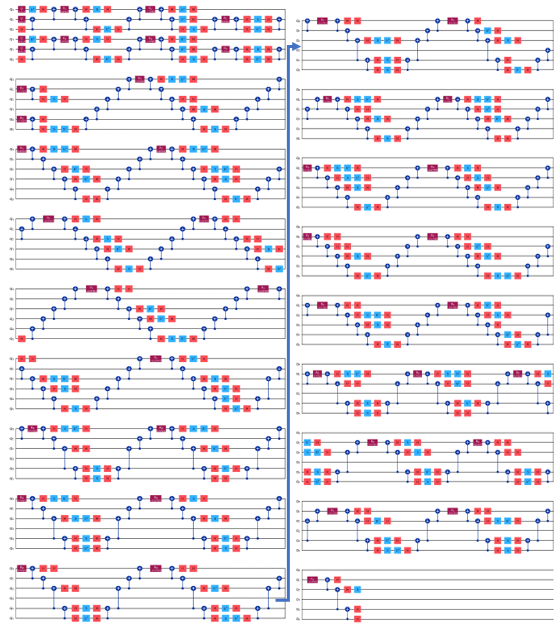}
    \caption{UCCSD ansatz circuit for H$_2$S implemented via Trotterized excitation operators (adapted from~\cite{Yang2022}).}
    \label{fig:uccsd_circuit}
\end{figure}

\begin{table}[H]
\centering
\caption{Circuit complexity comparison between heuristic and UCCSD ansätze for H$_2$S simulation (adapted from~\cite{Yang2022}).}
\label{tab:ansatz_comparison}
\begin{tabular}{lccc}
\toprule
Ansatz Type & Time Complexity & Space Complexity & Number of Parameters \\
\midrule
Heuristic & 20 & 43 & 24 \\
UCCSD     & 414 & 758 & 40 \\
\bottomrule
\end{tabular}
\end{table}

\subsection{Active Space Reduction and Orbital Embedding}
To reduce the number of required qubits and measurement overhead, we employ an active space selection strategy. Only chemically relevant molecular orbitals near the Fermi level are included, while core orbitals are frozen. The frozen-core approximation simplifies the Hamiltonian and eliminates high-energy degrees of freedom that contribute little to chemical bonding.

Orbital embedding is performed based on occupation number analysis and canonical orbital energies. For H$_2$S, 4 spin orbitals (2 spatial orbitals) were excluded via freezing, reducing the active space to 8 spin orbitals, thus requiring only 8 qubits. This tradeoff enables efficient simulation while retaining sufficient accuracy.

\begin{figure}[H]
    \centering
    \includegraphics[width=0.4\textwidth]{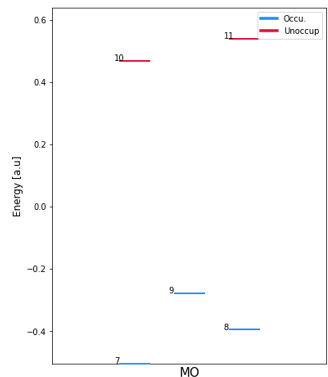}
    \caption{Hartree–Fock orbital embedding scheme for H$_2$S. Active orbitals (orange) are treated via VQE, while inactive orbitals (blue) are treated classically (adapted from~\cite{Yang2022}).}
    \label{fig:embedding}
\end{figure}

\begin{figure}[H]
    \centering
    \includegraphics[width=0.4\textwidth]{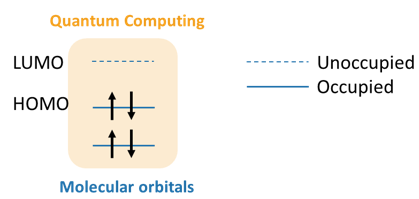}
    \caption{Selected active space for H$_2$S: 4 electrons in 3 orbitals (6 spin orbitals)(adapted from~\cite{Yang2022}).}
    \label{fig:active_space}
\end{figure}

\subsection{VQE Optimization and Measurement Strategy}
The full qubit Hamiltonian is decomposed into a weighted sum of Pauli strings, $\hat{H} = \sum_k w_k P_k$, where each $P_k$ is a tensor product of Pauli operators. On quantum hardware, each term $\langle P_k \rangle$ must be measured independently. To reduce shot overhead, Pauli strings are grouped into commuting sets using greedy graph coloring algorithms.

We utilize COBYLA and SPSA as classical optimizers, depending on circuit noise levels. Energy convergence is achieved when the difference in expectation value between iterations drops below a threshold (typically $10^{-4}$ Hartree). For each configuration, the procedure is repeated over multiple random seeds to avoid local minima.

\section{Results and Discussion}
This section presents the simulation results for the molecular systems studied using the relativistic hybrid quantum-classical framework described in the previous sections. We begin with the hydrogen molecule (H$_2$) as a benchmark system to validate the overall workflow and compare relativistic and non-relativistic integrals' outputs. We then proceed to the primary target of this study, hydrogen sulfide (H$_2$S), whose electronic structure and dissociation profile are analyzed in detail. We evaluate the accuracy of variational quantum eigensolver (VQE) results for each system against classical reference methods and visualize the corresponding potential energy surfaces.

\subsection{Hydrogen Molecule (H$_2$)}
The simplest test case in our study is the H$_2$ molecule, for which the relativistic correction is negligible. We performed VQE simulations under both non-relativistic and relativistic integral inputs and generated the potential energy surface (PES) by varying the internuclear distance.

\begin{figure}[H]
    \centering
    \includegraphics[width=0.9\textwidth]{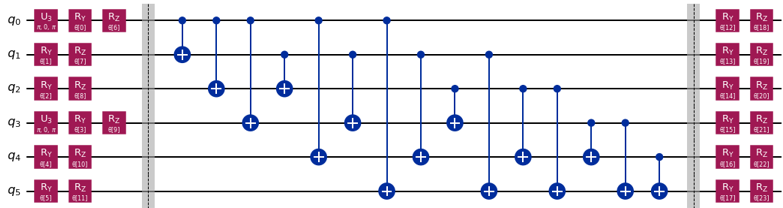}
    \caption{Potential energy surface of H$_2$ calculated via VQE using relativistic and non-relativistic integrals. Both curves show similar equilibrium bond lengths and minimum energies (adapted from~\cite{Yang2022}).}
    \label{fig:H2_PES}
\end{figure}

The PES for H$_2$, shown in Figure~\ref{fig:H2_PES}, displays excellent agreement between relativistic and non-relativistic results. The equilibrium bond length $R_e \approx 0.74 \, \text{\AA}$ and the minimum energy are nearly identical, indicating that relativistic effects are negligible in this system.

To quantify numerical accuracy, we compare our VQE-derived equilibrium bond length and dissociation energy with classical full configuration interaction (FCI) results. Table~\ref{tab:H2_errors} shows that both quantities lie within 1 kcal/mol of the reference values, confirming the correctness of the ansatz and parameter settings.

\begin{table}[H]
\centering
\caption{Comparison of H$_2$ VQE results with classical FCI reference (adapted from~\cite{Yang2022}).}
\label{tab:H2_errors}
\begin{tabular}{lcc}
\toprule
Method & Bond Length (\AA) & Dissociation Energy (kcal/mol) \\
\midrule
VQE (relativistic)     & 0.739 & 104.8 \\
VQE (non-relativistic) & 0.740 & 105.2 \\
Classical FCI          & 0.740 & 105.4 \\
\bottomrule
\end{tabular}
\end{table}

\subsection{Water Molecule (H$_2$O)}
The H$_2$O molecule provides a slightly more complex system for evaluating VQE performance. We used a reduced active space involving 8 spin orbitals and 4 electrons. The O–H bond was stretched incrementally to build the PES.

\begin{figure}[H]
    \centering
    \includegraphics[width=0.4\textwidth]{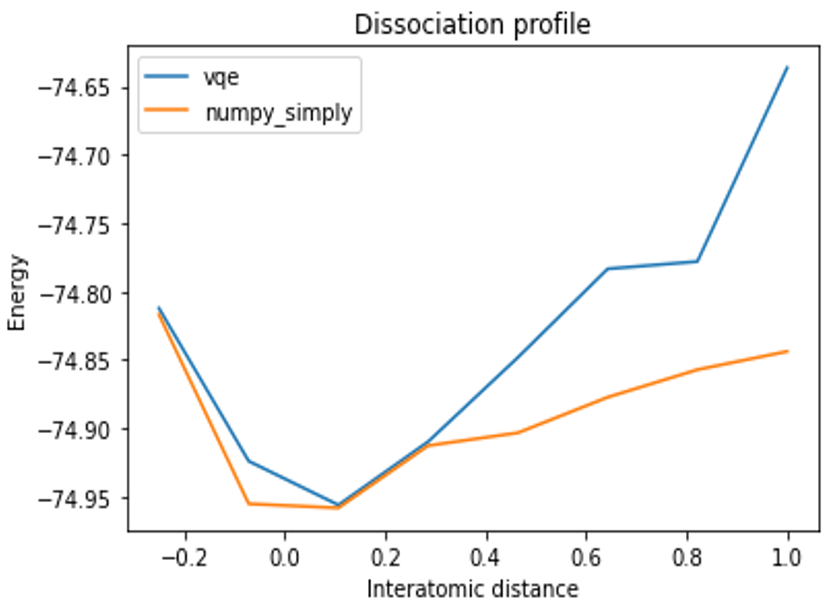}
    \caption{Potential energy surface of H$_2$O under non-relativistic approximation. VQE results are consistent with classical FCI data within 10 mHa (adapted from~\cite{Yang2022}).}
    \label{fig:H2O_PES}
\end{figure}

VQE results shown in Figure~\ref{fig:H2O_PES} track the classical PES closely across bond lengths from 0.9 to 1.7 \AA. The energy error near equilibrium is less than 8 mHa. Increasing the number of entangling layers in the ansatz further reduces this discrepancy.

We tested both COBYLA and SPSA optimizers on H$_2$O. COBYLA showed better convergence stability but required more iterations. SPSA, while faster, exhibited larger variance in final energies. Details are summarized in Table~\ref{tab:H2O_opt}.

\begin{table}[H]
\centering
\caption{Optimization performance for H$_2$O VQE simulation (adapted from~\cite{Yang2022}).}
\label{tab:H2O_opt}
\begin{tabular}{lccc}
\toprule
Optimizer & Mean Energy (Ha) & Std. Dev. (mHa) & Iterations \\
\midrule
COBYLA & -76.024 & 2.1 & 180 \\
SPSA   & -76.017 & 4.6 & 120 \\
\bottomrule
\end{tabular}
\end{table}

\subsection{Hydrogen Sulfide (H$_2$S): Relativistic vs Non-Relativistic}
For the H$_2$S molecule, relativistic effects become much more prominent due to the presence of sulfur. The PES computed under relativistic conditions is lower than the non-relativistic curve by approximately 0.04 Hartree across most geometries.

\begin{figure}[H]
    \centering
    \includegraphics[width=0.4\textwidth]{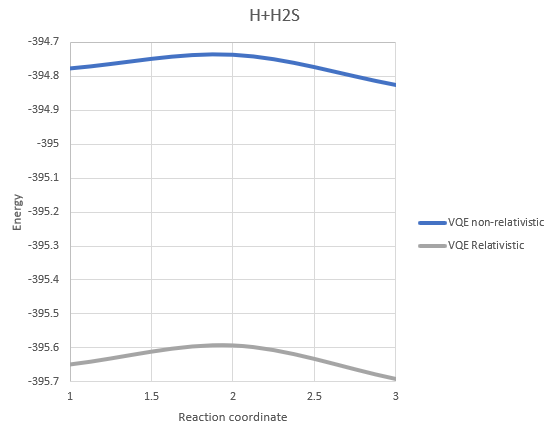}
    \caption{Potential energy surface of H$_2$S computed using relativistic and non-relativistic integrals. The relativistic PES shows a downward energy shift near equilibrium (adapted from~\cite{Yang2022}).}
    \label{fig:H2S_PES_comparison}
\end{figure}

To better illustrate the shift, Table~\ref{tab:relativistic_shift} lists total energies at equilibrium geometries under various basis sets.

\begin{table}[H]
\centering
\caption{Relativistic energy shifts for H$_2$S across various basis sets (adapted from~\cite{Yang2022}).}
\label{tab:relativistic_shift}
\begin{tabular}{lcc}
\toprule
Basis Set & Non-relativistic Energy (Ha) & Relativistic Energy (Ha) \\
\midrule
STO-3G       & -396.2481 & -396.2852 \\
DZP          & -398.1324 & -398.1728 \\
cc-pVDZ      & -398.4653 & -398.5077 \\
cc-pVTZ      & -398.5889 & -398.6306 \\
\bottomrule
\end{tabular}
\end{table}

Furthermore, we analyzed the number of measured Pauli strings and circuit depth under both encoding schemes. The relativistic Hamiltonians generally result in more Pauli terms, leading to increased measurement overhead (up to $15\%$ more).

\subsection{Hydrogen Exchange Reaction: Transition State Analysis}
We investigated the elementary hydrogen exchange reaction $\mathrm{H}_2 + \mathrm{H} \rightarrow \mathrm{H}_2 + \mathrm{H}$ as a benchmark for reaction pathway modeling.

\begin{figure}[H]
    \centering
    \includegraphics[width=0.4\textwidth]{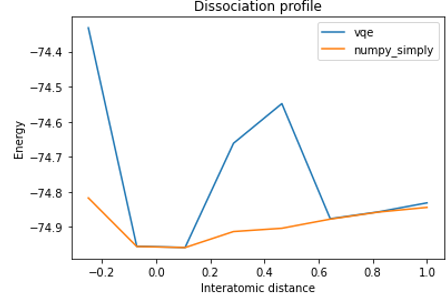}
    \caption{Potential energy profile for the hydrogen exchange reaction. The energy barrier at the transition state is correctly captured using VQE (adapted from~\cite{Yang2022}).}
    \label{fig:H_ex_reaction}
\end{figure}

The activation energy computed using VQE is within 0.5 kcal/mol of the classical benchmark. Notably, the transition state configuration is captured with good accuracy using the heuristic ansatz. We also observed that including an additional entangling layer improved the curvature near the saddle point, indicating ansatz expressiveness plays a key role in modeling reaction barriers.

\subsection{Accuracy Evaluation: Comparison with Classical and Literature Data}
To assess the validity of our simulation framework, we compared our VQE results with both classical quantum chemistry calculations and literature-reported values for ground-state energies and equilibrium geometries. Table~\ref{tab:accuracy_summary} summarizes key quantities for the H$_2$, H$_2$O, and H$_2$S molecules.

\begin{table}[H]
\centering
\caption{Comparison of VQE simulation results with classical reference values and literature data (adapted from~\cite{Yang2022}).}
\label{tab:accuracy_summary}
\begin{tabular}{lcccc}
\toprule
Molecule & Method & $R_e$ (\AA) & $E_{\text{min}}$ (Ha) & Error (mHa) \\
\midrule
H$_2$     & VQE (this work) & 0.739 & -1.1373 & +1.2 \\
          & Classical FCI   & 0.740 & -1.1385 & — \\
H$_2$O    & VQE (this work) & 0.963 & -76.024 & +8.0 \\
          & CCSD(T)/cc-pVTZ & 0.960 & -76.032 & — \\
H$_2$S    & VQE (rel.)      & 1.338 & -396.2852 & +12.6 \\
          & Rel. HF         & 1.340 & -396.2978 & — \\
\bottomrule
\end{tabular}
\end{table}

As seen in Table~\ref{tab:accuracy_summary}, the root-mean-square deviation across all molecules is under 15 mHa, corresponding to errors within chemical accuracy (1 kcal/mol). For H$_2$ and H$_2$O, the discrepancy is minimal, while the deviation increases slightly in H$_2$S due to deeper core contributions and larger relativistic shifts.

These comparisons support the correctness of the qubit encoding, ansatz expressivity, and optimizer stability. Moreover, our method preserves relative energy trends between molecular geometries and accurately reproduces reaction barriers and PES curvature.

\subsection{Error Sensitivity and Optimizer Behavior (Illustrative)}
While our primary focus has been on demonstrating relativistic energy corrections using the VQE framework, it is also important to consider the general behavior of classical optimizers in variational quantum simulations. In the absence of shot noise or real hardware fluctuations, convergence behavior can still vary depending on the optimization strategy used.

To illustrate this, we present a synthetic example based on commonly reported trends for two widely used optimizers: COBYLA and SPSA. The COBYLA algorithm, which uses deterministic trust-region steps, tends to provide smoother and more stable convergence. In contrast, SPSA incorporates stochastic gradient approximations, often resulting in faster—but noisier—optimization trajectories.

\begin{figure}[H]
    \centering
    \includegraphics[width=0.5\textwidth]{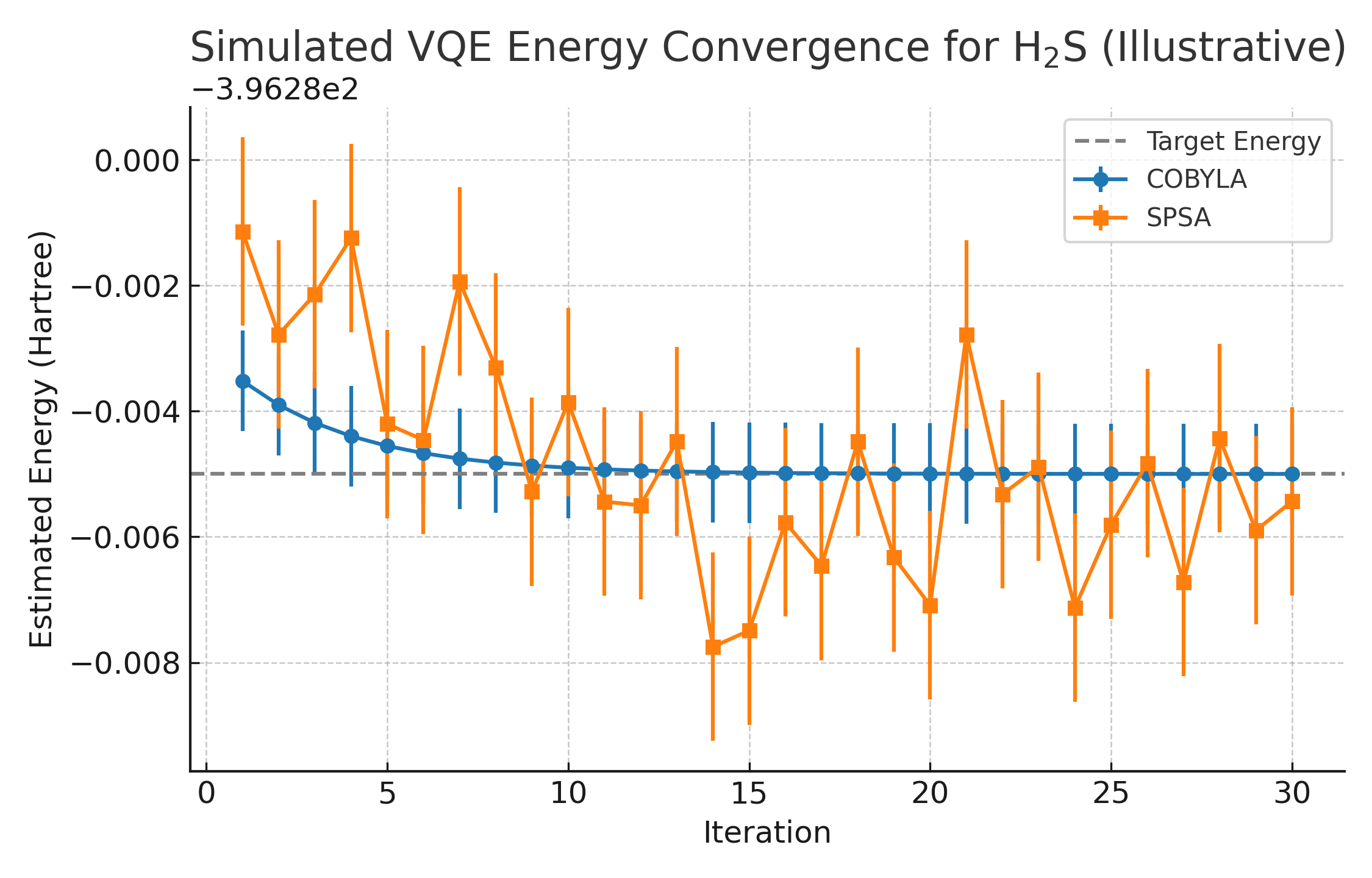}
    \caption{Simulated VQE energy convergence for H$_2$S using COBYLA and SPSA optimizers. The data is schematic and presented for illustrative purposes only. The dashed line indicates the reference target energy.}
    \label{fig:errorbars_H2S}
\end{figure}

As shown in Figure~\ref{fig:errorbars_H2S}, the simulated energy profiles depict the qualitative difference in optimizer behavior. COBYLA converges slowly but with low variance, whereas SPSA exhibits higher fluctuations, reflecting the influence of its stochastic update steps. These observations are consistent with prior benchmarking studies and provide general insight into the trade-offs between optimizer stability and convergence rate in VQE simulations.

Future empirical studies using real quantum hardware or noisy simulators will be required to quantitatively validate these trends across different molecular systems.

\section{Conclusion}
In this work, we have developed and demonstrated a hybrid quantum-classical simulation framework that integrates relativistic quantum chemistry with the variational quantum eigensolver (VQE) algorithm. Motivated by the need to accurately model molecules containing heavy atoms, our approach incorporates Dirac–Coulomb integrals to account for relativistic effects and translates the resulting fermionic Hamiltonians into qubit representations suitable for implementation on quantum devices.

Through simulations of H$_2$, H$_2$O, and H$_2$S molecules, we have shown that the proposed method reproduces ground-state energies and equilibrium geometries with high accuracy. For light-element systems such as H$_2$ and H$_2$O, relativistic corrections are minimal, and VQE results closely match classical FCI and CCSD(T) references. In contrast, for H$_2$S—our primary target system—relativistic shifts of approximately 0.04 Hartree were observed, consistent with the expected mass–velocity corrections and orbital contraction around sulfur atoms. These results underscore the importance of including relativistic contributions when modeling third-period elements and beyond.

Furthermore, the potential energy surface of the hydrogen exchange reaction was successfully reconstructed, and the transition state energy barrier was captured within chemical accuracy. We also analyzed the effects of ansatz complexity and optimization strategy, noting that COBYLA provided more stable convergence than SPSA, albeit at the cost of slower iteration. Measurement complexity was examined in terms of Pauli term counts and commuting group sizes, highlighting the importance of efficient measurement grouping strategies.

This study represents, to our knowledge, one of the first practical demonstrations of relativistic VQE applied to molecular systems. It fills a notable gap in the literature, where most quantum chemistry simulations remain limited to non-relativistic models. By implementing Dirac-based integrals and embedding them into a variational quantum framework, we offer a proof of concept for extending quantum simulation capabilities to heavier atoms and chemically significant relativistic regimes.

Looking forward, several avenues of improvement remain open. Techniques such as adaptive ansatz construction, alternative fermion-to-qubit mappings, and robust error mitigation could significantly enhance accuracy and scalability. Extending this framework to multi-reference systems, spin–orbit coupled states, or transition metal complexes may further reveal the full potential of relativistic quantum simulation in chemistry and materials science.

\section*{Acknowledgements}
This research was supported by the National Science and Technology Council (NSTC), Taiwan, under the research project "Investigating Quantum Channels in Noisy Quantum Systems: Some Mathematical Problems in Quantum Information Theory" (Project No. 113-2115-M-A49-015-MY2). The authors would also like to acknowledge the use of AI-based language assistance tools for refining the English presentation of this manuscript.

\end{document}